\documentclass[sigconf]{acmart}

\copyrightyear{2024}
\acmYear{2024}
\setcopyright{acmlicensed}
\acmConference[IMC '24] {Proceedings of the 2024 ACM Internet Measurement Conference}{November 4--6, 2024}{Madrid, Spain.}
\acmBooktitle{Proceedings of the 2024 ACM Internet Measurement Conference (IMC '24), November 4--6, 2024, Madrid, Spain}
\acmISBN{979-8-4007-0592-2/24/11}
\acmDOI{10.1145/3646547.3688422}

\usepackage[english]{babel}
\usepackage{blindtext}
\usepackage{graphicx}
\usepackage{subcaption}
\usepackage{bm}
\usepackage{cleveref}

\usepackage[medium,compact]{titlesec}
\usepackage{titlesec}
\titleformat{\section}
  {\normalfont\Large\bfseries\MakeUppercase}{\thesection}{1em}{}

\newcommand{\parb}[1]{\vspace{1pt}\noindent{\bfseries #1.}}
\newcommand{\parbi}[1]{\vspace{1pt}\noindent{\bfseries\itshape #1.}}
\newcommand{\boldbox}[1]{\vspace{8pt}\hrule height 0.75pt\nopagebreak\vspace{2pt}\noindent\textbf{#1}\vspace{2pt}\nopagebreak\hrule height 0.75pt\vspace{4pt}}

\usepackage{url}

\usepackage{breakurl}

\usepackage{xspace}
\newcommand{\ie}{\emph{i.e.,}\xspace}
\newcommand{\eg}{\emph{e.g.,}\xspace}
\newcommand{\etal}{\emph{et al.}\xspace}

\usepackage{enumitem}
\usepackage{multirow}
\usepackage{hyperref}

\def\Snospace~{\S{}}

\newcommand{\name}{CPT-GPT\xspace}

\newcommand{\jonny}[1]{{\em \color{magenta} (Jonny: #1)}}

\newcommand{\cut}[1]{{}}
\begin{document}
\title[]{High-Fidelity Cellular Network Control-Plane Traffic Generation
  without Domain Knowledge}




\author{Z. Jonny Kong}
\affiliation{
  \institution{Purdue University}
  \city{West Lafayette}
  \state{IN}
  \country{USA}
}
\author{Nathan Hu}
\affiliation{
  \institution{Stanford University}
  \city{Palo Alto}
  \state{CA}
  \country{USA}
}
\author{Y. Charlie Hu}
\affiliation{
  \institution{Purdue University}
  \city{West Lafayette}
  \state{IN}
  \country{USA}
}
\author{Jiayi Meng}
\affiliation{
  \institution{University of Texas at Arlington}
  \city{Arlington}
  \state{TX}
  \country{USA}
}
\author{Yaron Koral}
\affiliation{
  \institution{AT\&T Labs}
  \city{Middletown}
  \state{NJ}
  \country{USA}
}


\begin{abstract}

With rapid evolution of mobile core network (MCN) architectures,
large-scale control-plane traffic (CPT) traces are critical to studying 
MCN design and performance optimization by the R\&D community.
The prior-art control-plane traffic generator SMM heavily relies
on domain knowledge which requires re-design as the domain evolves.
In this work, we study the feasibility
of developing a high-fidelity MCN control plane traffic generator
by leveraging generative ML models.
We identify key challenges in synthesizing high-fidelity CPT
including generic (to data-plane) requirements such as multi-modality feature relationships
and unique requirements such as stateful semantics and long-term (time-of-day) data variations.
We show state-of-the-art, generative adversarial network (GAN)-based
approaches shown to work well for data-plane traffic
cannot meet these fidelity requirements of CPT,
%
and develop a transformer-based model, \name,
that accurately captures complex dependencies among the samples in each traffic stream (control
events by the same UE) without the need for GAN.
Our evaluation of  \name on a large-scale control-plane traffic
trace shows that
(1) it does not rely on domain knowledge yet
synthesizes control-plane traffic with comparable fidelity as SMM;
(2) compared to the prior-art GAN-based approach, it
reduces the fraction of streams that violate stateful semantics by
two orders of magnitude,
{
the max y-distance of sojourn time distributions of streams by 16.0\%, 
and the transfer learning time in deriving new hourly models by 3.36$\times$.
}
\end{abstract}

\maketitle

\section{Introduction}
\label{sec:intro}

The recently introduced Control-/User-Plane Separation (CUPS) in 3GPP
Release~14 for 4G~\cite{3gpp.23.214} and in 3GPP Release~15
for 5G~\cite{3gpp.23.501},
combined with a significant increase in
control-plane traffic alongside explosive growth in data-plane
traffic~\cite{nokia,control_plane_growth}, challenge mobile network
operators and designers to innovate on mobile network architectural
design not only for the data plane but also for the control plane 
in order to provide sustained mobile user experience.

Indeed, an increasing number of efforts have focused on open-source
development of 3GPP-compliant MCN implementations~\cite{free5gc} and
novel MCN designs for improved
performance~\cite{jain2022l25gc,ferguson2023corekube,hasan2023building}.
Accurately assessing the performance of such open-source
implementations or new MCN designs under real deployment scenarios,
however, critically relies on using high-fidelity, large-scale control-plane
traffic to drive the MCN operations.
Despite the need, large-scale control-plane traffic traces in MCN
are only accessible by mobile network operators, who are unlikely
to share their traffic traces due to business and privacy concerns.
As a result, the lack of public MCN control-plane traffic
hinders the in-depth study of MCN design and performance optimization
by the broad networking and systems communities.

A classic approach to mitigating the lack of
real network traffic traces is to generate synthetic traces.
However, the large body of traffic synthesis work
for LTE and 5G has focused on the data plane of MCN, 
\ie on modeling data-plane traffic 
(\eg ~\cite{lindberger1999balancing,li2010dimensioning,li2010dimensioning2,checko2012capacity,jailani2012lte}).
%
On the control-plane side,
Dababneh~\etal~\cite{dababneh2015data} modeled the total control-plane volume
on LTE's MCN but did not model the fine-grained inter-arrival time of
successive events or the intricate event dependence specified by 3GPP for each
individual UE; they also ignored the traffic diversity in device types and
time-of-day.


Compared to Internet data traffic generation, cellular network control-plane
traffic generation faces some common fidelity requirements but also unique new
requirements.
A control-plane traffic trace consists of multiple streams of control events
(samples), one stream per UE, typically throughout a day.
{\bf (1) Stateful semantics}. First and foremost, the traffic generator must accurately capture the
rigid inter-dependence of sample features within a stream, adhering to
domain-specific rules, \ie 3GPP UE state machines, ensuring only semantically
correct generated streams will be used to drive the MCN.
{\bf (2) Multi-modal features}. Like data traffic, control-plane traffic
encompasses multiple fields.
Specifically, each sample includes a control event type and a timestamp, and
thus the traffic generator is tasked not only with producing realistic
distributions for individual fields, but also generating realistic correlations
between fields within each sample.
{\bf (3) Variable flow length}. As in data traffic, the traffic generator needs
to capture diversity in UE flow length.
For instance, the number of events within a fixed time window could vary among
clients, which follows from the diversity in UE activity levels.
{\bf (4) Long-term data drifts}. As control-plane traffic lasts throughout the
day, the traffic generator needs to capture data distribution drifts that occur
over time, such as variations throughout a day.

To meet these requirements, the state-of-the-art cellular control-plane traffic
generator SMM~\cite{meng2023modeling} employs a traditional modeling methodology.
In particular, it derives a statistical model, \ie a Semi-Markov model, based
on  the 3GPP UE state machines and fits the model parameters (state transitions
probabilities and sojourn times) on the real trace.
Such a traditional modeling approach, however, critically relies on domain
knowledge, \eg the 3GPP standard, and hence requires re-design as the domain
knowledge evolves.
First, a separate model needs to be designed for each generation of
cellular technology (4G or 5G).
%
Second, even for a particular generation of technology, standards evolve
continuously over time as new 3GPP specifications are published. For example,
the development of LTE was carried out through 3GPP Release 8 (2008) to Release
16 (2020)~\cite{about3gpp}.
Therefore, continuous manual effort is required to keep such
domain-knowledge-dependent generators up to date with the latest standards.

An additional limitation of the above traditional modeling approach is that the
model parameters are rigid, and a single model cannot capture the diversity of
control-plane traffic trace. 
As a result,
the authors had to cluster
control-plane-traffic and instantiate 20,216 models
(one per cluster per hour) which are inconvenient to maintain and deploy.

In light of the limitation of the traditional modeling approach,
in this work, we study the feasibility
of developing a high-fidelity MCN control plane traffic generator
by leveraging generative machine learning (ML) models.
%
Our approach was inspired by the recent work on using  Generative
Adversarial Networks (GAN) combined with LSTM in synthesizing {\em data-plane traffic}
traces~\cite{doppelgan:imc2020,netshare:sigcomm2022} which showed
GAN-based approaches effectively learn
the distributions and correlations of packet- or flow-level header fields in Internet traffic.

However, when we adapted prior-art GAN-based approaches 
to generate {\em control-plane traffic} of the cellular network, 
we found it suffers a number of limitations.
(1) The prior-art approaches such as
NetShare~\cite{doppelgan:imc2020}
cannot capture the stateful semantics,
as 22.10\% of the streams in the synthesized trace contain one or more events
that violate state transition rules stipulated by the 3GPP protocols.
(2) NetShare fails to capture fine-grained temporal properties, such as the
distribution of sojourn times, \ie the duration that a UE stays in a
particular state of the 3GPP state machine.
{
For example, the maximum y-distance between the CDFs of the real and
synthesized traces' sojourn times reaches 61.7\%.
}
(3) The GAN-based approach cannot efficiently
adapt to data distribution drifts that occur through time via transfer
learning.
In particular, training a model for one hour and applying
transfer learning for each of the next 5 hours takes almost
2X as long as direct training a 6-hour model from scratch.
(4) These GAN-based LSTM approaches require complex and specialized
enhancements to alleviate mode collapse~\cite{kodali2017convergence}.
(5) Finally, these approaches require complex enhancements to the
model design to address LSTM's tendency to forget past
states~\cite{schak2019study,arora2019does} in order to capture
long-term dependencies.

%
To overcome these drawbacks, we develop a transformer-based
framework, \name,
that accurately captures complex dependencies among the events in each 
stream to meet the fidelity requirements (stateful semantics and distribution of multiple
features).
In doing so, it not only achieves comparable fidelity
in traffic generation without domain knowledge as needed in SMM,
but also achieves higher fidelity than
prior-art GAN/LSTM-based approaches~\cite{doppelgan:imc2020,netshare:sigcomm2022}
without the need to use GAN or
%
model enhancements such as dealing with long-term dependencies and mode collapse.
Instead, the design of \name only has to focus on a few challenges in the input
layer and output layer design:
{
{(1) How to design tokens for the input layer to capture multi-modal features
per sample?}
%
%
{(2) How to introduce generation stochasticity at the output layer for
multi-modal features?} %
}
%
{(3) How to efficiently update the model over time?}
%

We implement the complete cellular network control-plane generation model
\name in PyTorch, based on a decoder-based transformer
model~\cite{vaswani2017attention}.
We extensively evaluate \name by comparing it with the traditional
modeling-based SMM~\cite{meng2023modeling}
and the SOTA GAN-based
NetShare~\cite{netshare:sigcomm2022}, on a large-scale control-plane traffic trace
(for 380K UEs) from a leading mobile operator.
Our evaluation shows that 
(1) Compared to SMM,
\name requires no domain knowledge, yet it can achieve close-to-zero semantic
violations, similarly accurate sojourn time distribution in terms of the max
y-distance of the CDFs, {and more accurate percentage breakdown of
different event types.}
For example, averaged over the three types of UEs (phones,
connected cars, and tablets), \name's differences compared to the real dataset
in the breakdown of the two dominant event types (\texttt{SRV\_REQ} and
\texttt{S1\_CONN\_REL}) are 1.01\% and 1.21\% lower than SMM, respectively.
%
(2) Compared to NetShare, \name achieves much higher fidelity in
      control-plane traffic generation.
      (i) It
reduces the fraction of streams that violate stateful semantics by two
orders of magnitude, from 22.1\%, 11.5\%, and 16.9\% to 0.2\%, 0.4\%, and 1.5\%
for phones, connected cars, and tablets, respectively.
(ii) It also improves the sojourn time distribution of
the synthesized traces.
For the three device types, the max y-distance of sojourn time distribution
from the real trace is reduced by 10.7\%, 9.1\%, and 28.3\% respectively,
averaging over the two 3GPP states.
%
%
%
(iii) \name is more efficient in being applied transfer learning to adapt a
pretrained model for a given hour-of-day to the subsequent hours. Specifically,
it reduces the time needed to derive hourly models by 3.36$\times$.
(3) \name generates diverse control-plane traffic without memorizing individual
sequences from the training set.
Specifically, when examining generated sub-sequences of length 20, none are
repeated from the training set.

  In summary, this work makes the following contributions:
\begin{itemize}[nosep,left=0pt]%
\item We develop the first control-plane traffic generator that does not
    require domain knowledge yet achieves comparable accuracy as the prior-art
    Semi-Markov-Model-based traffic generator which heavily relies on domain
    knowledge.
\item We show how transformer-based models can overcome the low fidelity
    limitations of prior-art GAN/LSTM-based models in synthesizing
    control-plane traffic, more efficiently employ transfer learning to deal
    with traffic shifts over time, and does not memorize
    individual sequences from the training set.
%
%
\end{itemize}

\section{Background \& Motivation}
\label{sec:moti}

\subsection{Cellular Network Control-Plane Traffic}
The Mobile Core Network (MCN) serves as the central hub of the cellular network. 
It not only forwards data traffic between each user equipment (UE) and the Internet,
but also orchestrates various functions of the cellular network 
to provide seamless and reliable communication services to the users.
To effectively manage both data and control traffic within the cellular
network, Control-/User-Plane Separation (CUPS) was first introduced in 3GPP
Release~14 for 4G~\cite{3gpp.23.214} and further refined in 3GPP Release~15 
for 5G~\cite{3gpp.23.501}.
This separation divides the cellular network into a data plane and a control plane,
streamlining the processing of data and control traffic, respectively.

\begin{table}[tp]
    \centering
    \caption{Control-plane event types in 4G and 5G.}
   \label{tab:control_event}
{\small
    \begin{tabular}{c|c|p{4cm}}
      \toprule
      \textbf{4G} & \textbf{5G} & \textbf{Description} \\
       \midrule
      \texttt{ATCH} & \texttt{REGISTER} 
       & Register the UE with the MCN \\
      \texttt{DTCH} & \texttt{DEREGISTER} 
       & De-register the UE from the MCN \\
      \texttt{SRV\_REQ} & \texttt{SRV\_REQ} 
       & Create a signaling connection to allow UE to send/receive data and control-plane messages\\
      \texttt{S1\_CONN\_REL} & \texttt{AN\_REL} 
       & Release the signaling connection and others resources in both control and data planes \\
      \texttt{HO} & \texttt{HO}
       & Switch the UE from the current cell coverage serving it to another cell \\
      \texttt{TAU} & $-$ 
       & Update the UE's tracking area \\
      \bottomrule
   \end{tabular}
}
\end{table}

\noindent
{\bf Control events.}
To effectively manage and control the communications in the cellular network, 
3GPP defines various types of control-plane events for both LTE and 5G.

\Cref{tab:control_event} lists the 
primary control-plane events in 4G and
5G and 
describes their main functionalities.
Note that (1) we ignore the events that happen only between the UE and the
Radio Access Network (RAN), as they do not involve the MCN;
(2) although each event corresponds to a series of control-plane
  messages among UE, RAN, and MCN, traffic synthesis
  is only concerned with generating control events originated from UEs,
  as mapping from a control-plane event to messages is fixed as dictated
  by the 3GPP protocol;
  and (3) what to do in case of event failure, \ie due to message
  exchange failure, is the responsibility of the downstream
  applications, which is beyond the scope of this paper; the
  traffic generator is concerned with generating semantically correct
  sequences of events.

\begin{figure*}
\centering
    \begin{subfigure}{.6\textwidth}
        \centering
        \includegraphics[width=.99\linewidth]{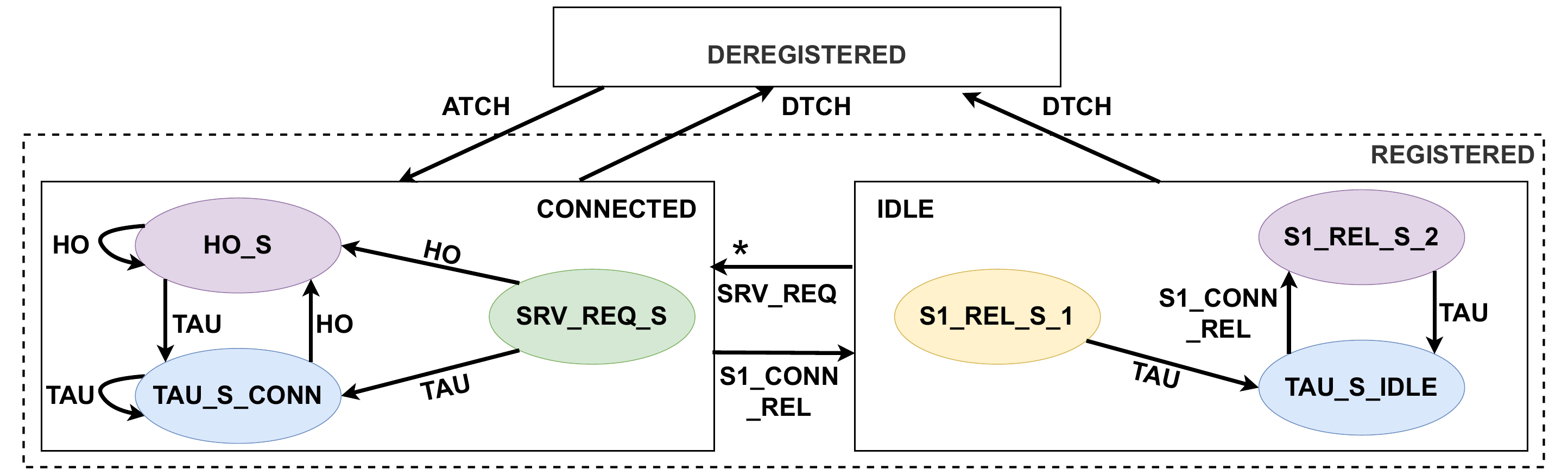}
        \caption{4G}
        \label{fig:lte_sm}
    \end{subfigure}%
    \hfill
    \begin{subfigure}{.4\textwidth}
        \centering
        \includegraphics[width=.99\linewidth]{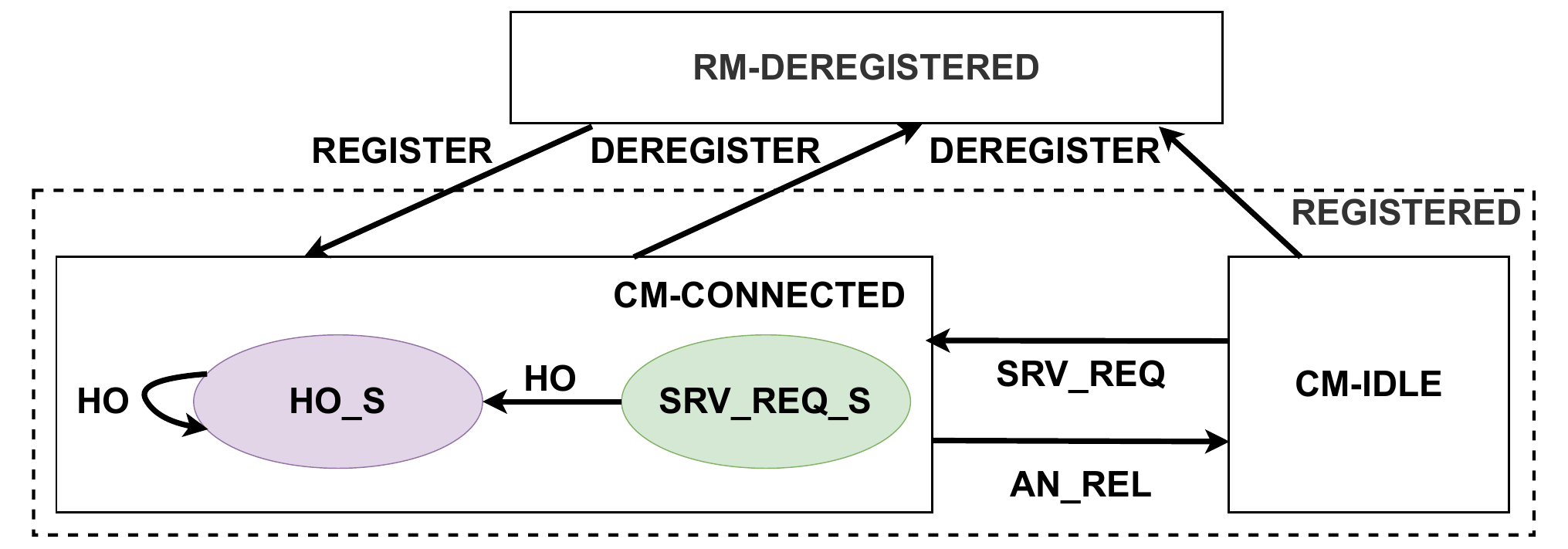}
        \caption{5G}
        \label{fig:5g_sm}
    \end{subfigure}
\caption{The two-level hierarchical UE state machines of 4G and 5G~\cite{meng2023modeling}.}
\label{fig:lte_5g_sm}
\end{figure*}

\noindent
{\bf Stateful semantics of control-plane traffic.}
In contrast with the stateless Internet traffic, 
the control-plane events in the cellular network are not independent. 
The 3GPP protocol not only specifies that 
every UE has to follow two state machines 
when interacting with the MCN,
but also describes intricate dependence of the control events on the states in the
UE state machines.
Specifically, the two state machines for 4G/5G are
EMM/RM (EPS\footnote{EPS: Evolved Packet System, comprising the RAN and the MCN
in 4G.} Mobility Mgmt. / Registration Mgmt.) and ECM/CM (EPS Conn.~Mgmt. /
Conn.~Mgmt.)
respectively~\cite{3gpp.23.401, 3gpp.23.501}, each with two primary states
(\texttt{REGISTERED} and \texttt{DEREGISTERED} for EMM/RM, and
\texttt{CONNECTED} and \texttt{IDLE} for ECM/CM).

%

To capture the intricate dependence of the control events on the above
four UE states, \cite{meng2023modeling} developed two-level
hierarchical state machines for 4G and 5G as shown in
\autoref{fig:lte_5g_sm}.
\autoref{fig:lte_sm} shows the two-level state machine of 4G.
The top-level state machine is a merged state machine of 
the EMM and ECM state machines with three UE states, 
\texttt{DEREGISTERED}, \texttt{CONNECTED}, and \texttt{IDLE},
derived from insights into how a UE transitions in the two state machines.
%
At the bottom level,
there are two sub-state machines 
embedded in the top-level \texttt{CONNECTED} and \texttt{IDLE} states,
which capture other dependence of the control events
on the top-level states.

While 4G and 5G exhibit similarities in control events and UE states,
it is necessary to customize and re-derive the two-level state machine of 5G
to align with the 3GPP protocol as shown in \autoref{fig:5g_sm}.
Specifically, as \texttt{TAU} is not supported in 5G,
the corresponding states and transitions are removed 
from the two-level state machine of 4G.
Moreover, several 4G event types,
\texttt{ATCH}, \texttt{DTCH}, \texttt{S1\_CONN\_REL}, are replaced with 
\texttt{REGISTER}, \texttt{DEREGISTER}, and \texttt{AN\_REL},
respectively.

\subsection{Motivation for Generating Control Traffic}



%
To provide sustained mobile user experience,
mobile network operators and designers continuously innovate on mobile network
architectural design not only for the data plane but also for the
control plane.
Such innovations 
rely on realistic, high-fidelity control-plane traffic traces
for evaluation, benchmarking, and optimization.
We describe two motivating use cases in data-driven network
design and management that demand realistic control-plane traffic
traces.

\textbf{Performance evaluation of MCN design.}
%
In-depth evaluation of various aspects of novel MCN design
(\eg~\cite{jain2022l25gc,ferguson2023corekube,hasan2023building}), including
throughput, end-to-end latency, scalability, and fault resilience,
in handling
control events originating from a large UE population rely on realistic 
control-plane workload.
For example, right after the first control-plane traffic
generator~\cite{meng2023modeling} became available, the Aether community
started using it to study the scalability of Aether 5G core design.
%

\textbf{Real-time network management.}  
Network management is critical to various stakeholders ranging from network operators to end users,
It has been intensively studied for data flows (\eg five-tuples) via 
telemetry~\cite{phaal2001rfc3176, netflow,cormode2009count, charikar2002finding, yu2013software,
  liu2016one, huang2017sketchvisor, yang2018elastic, liu2019nitrosketch,
  agarwal2022heterosketch} 
which critically relies on real-time monitoring of network traffic. 
Accurate control-plane traffic models can
help to develop effective monitoring schemes, \eg with better accuracy and
smaller memory footprints.  For example, 
such models can help to determine
a good sampling rate for sampling-based monitoring
when collecting telemetry metrics.


\section{Overview}
\label{sec:overview}

We formally state the control-plane traffic generation problem and discuss
design challenges and how the prior-art heavily relies on domain knowledge.

\subsection{Problem Formulation}


A cellular control traffic dataset $D=\{S_1, S_2, ..., S_n\}$ consists of
multiple streams, where each stream $S_i$ represents a sequence of events
generated by a specific UE.
Each stream $S_i=\{UE\_ID_i, \allowbreak device\_type_i, \allowbreak
events_i\}$ consists of a UE identifier, a UE device type, and a sequence of
events, $events_i=\{\{t_{i1}, e_{i1}\}, \allowbreak \{t_{i2}, e_{i2}\},
\allowbreak ..., \allowbreak \{t_{ik}, e_{ik}\}\}$, where each event $e_{ik}$
is accompanied by a timestamp $t_{ik}$ indicating when it occurred.

Given such a control plane traffic dataset $D$, our objective is to develop a
framework
capable of producing a synthesized dataset $D'$ of arbitrary size that exhibits
high fidelity in terms of a selection of fidelity metrics of importance to the
downstream use cases.

\subsection{Challenges}
\label{subsec:overview_challenges}

The primary challenges in synthesizing high-fidelity control-plane traffic
datasets come from the generality of such a framework
and
from the multiple dimensions of fidelity required by the use cases
of control-plane datasets:

\begin{itemize}[leftmargin=*]
\item C1: \textit{Generality -- requiring minimal domain
  knowledge}. The framework should require as little domain
  knowledge as possible, since relying on domain knowledge leads to extra
  re-design effort upon domain knowledge changes, \eg revisions or new releases
  of 3GPP standards, or mixed technology deployment scenarios.
  Furthermore, 
  even within the same generation of network technology, the standard evolves
  continuously with new 3GPP releases.
  For example, LTE has been developed through 3GPP Release 8 to
  16~\cite{about3gpp}, whereas 5G began with 3GPP Release 15 and still actively
  evolving~\cite{ghosh20195g}.
  As the network standard continuously evolves, 
  the control-plane traffic of UEs experiences a continuous shift which requires
  the models to be re-instantiated repeatedly.
\item C2: \textit{Stateful semantics}.
  Each stream of the control-plane traffic trace consists of samples, \ie control events, each
  of a particular event type (Table~\ref{tab:control_event}).
  The framework needs to capture
  the rigid inter-dependence of event types, dictated by
  domain rules (\eg 3GPP UE state machines). Semantic correctness is
  essential since only semantically correct datasets shall
  be used, \eg in comparing different designs or implementations of an MCN.
\item C3: \textit{Multimodal features}. 
  In addition to event types, the samples in each control-plane stream also
  encompass several other fields, including
  an   associated timestamp, that directly affect the load on the MCN.
  Thus, the framework needs to generate control-plane traffic with realistic distributions for each field,
  which are required to evaluate how an MCN operates under
  real-world workloads, such as the duration an MCN has to maintain per-UE
  states in stateful implementations (\eg~\cite{ferguson2023corekube}).
  These include distributions (breakdowns) of different control event types in
  the trace and the sojourn time of a UE staying in each state,
  as each type of events will invoke a different set of network functions
  in the MCN, and the sojourn time staying in each UE state
  affects 
  the average and bursty load incurred by the UEs on the MCN.
\item C4: \textit{Variable flow length}.
Like Internet data traffic,
the framework needs to generate datasets
    consisting of streams of varying lengths, with the length distribution matching that
    of the real datasets. Realistic flow length distribution is important
    because it reflects the realistic workload experienced by the MCN
    in real deployment due to individual streams (UEs).

\item C5: \textit{Long-term data drifts}. The framework needs to
  capture control-plane traffic drifts over time, such as diurnal variations,
  due to variations of UE activity characteristics throughout different hours of
  the day. Accurately modeling control-plane traffic drifts enables
  {evaluating autoscaling capabilities of MCN
  implementations~\cite{ferguson2023corekube,hasan2023building}. }
\end{itemize}

%

\begin{table*}[tp]
\caption{Fidelity metrics used for evaluating the synthesized dataset.}
\label{tab:metrics}
\resizebox{0.85\textwidth}{!}{
\begin{tabular}{l|l|l}
\toprule
\multicolumn{1}{c|}{Metric} & \multicolumn{1}{c|}{Definition} & \multicolumn{1}{c}{Goal} \\ \midrule
Semantic violation & Percentage of events and streams that violates 3GPP-defined state transitions & Evaluate C2 \\
Sojourn time distribution & The average time a client stays in a particular state & Evaluate C3 \\
Event type breakdown & Percentage of each event type & Evaluate C3 \\
Flow length distribution & Number of events in each stream & Evaluate C4 \\
Adaptability to data drifts & Training time needed to adapt to another dataset of different characteristics & Evaluate C5 \\ \bottomrule
\end{tabular}
}
\end{table*}

\noindent
{\bf Fidelity metrics.}
To quantify how well a framework tackles the above challenges (C2-C5)
in generating high-fidelity control-plane traffic, we use the list of
fidelity metrics shown in \autoref{tab:metrics}.
First, to calculate the percentage of {\em semantic violations} and {\em sojourn time
distributions}, we replay the synthesized trace against the 3GPP-dictated state
machine (\autoref{fig:lte_5g_sm}), and record the duration that the UE stays in
each state and the number of events that violate state transitions,
respectively.
Second, {\em event type breakdown} and {\em flow
length distribution} can be directly derived from the synthesized trace.
Finally, to evaluate the {\em adaptability to data drifts}, we assume a transfer
learning setup, where the model in the framework is trained on a trace
collected in a particular hour, and subsequently fine-tuned to a trace
collected from another hour that may have data distribution shift, and measure
the re-training time and accuracy of the adapted model.

\subsection{Prior-art Control-plane Traffic Generation}
\label{subsec:smm}

The state-of-the-art control-plane traffic generator
SMM~\cite{meng2023modeling} relies heavily on domain-specific
knowledge.
(1) To satisfy stateful semantics (C2),
the authors manually derived the above two-level 
hierarchical state machines for 4G and 5G (\autoref{fig:lte_5g_sm})
based on 3GPP standard
to capture the intricate inter-dependence of the control events on the
four UE states.
(2) They then manually converted the two-level state machine
into a Semi-Markov Model (SMM) to model 
the probability of the transition from one state to another and
the duration of a UE staying in one UE state before switching to another
(referred to as SMM sojourn time).
(3) In fitting the parameters of the SMM (to satisfy C3 and C4),
the authors applied further domain knowledge.
They discovered high diversity of control-plane traffic in terms of
device types, time-of-the-day, and across individual UEs,
and then clustered the UEs in the real trace
into hundreds of UE clusters, each with similar features such
as flow length and variation of sojourn time which are domain-specific,
and then instantiated an SMM for every UE cluster
for each device type and hour-of-day.
(4) To finally model sojourn time - a key SMM model parameter (to satisfy C3),
they again applied domain knowledge; they derived one CDF model for each transition in the SMM,
after discovering that traditional probability distributions 
used for modeling the inter-arrival time of Internet traffic
(\eg Poisson, Pareto, Weibull, and TCPlib) 
cannot accurately model the control-plane traffic in cellular networks.
As a result, a total of 20,216 Semi-Markov models were
derived to cover all 24 hours of the day and 
three device types,
\ie phones, connected cars, and tablets,
which required deriving a total of 283,024 CDFs to model the sojourn time
in all the models.

As SMM's design heavily relies on domain knowledge,
\eg the 3GPP standard, it needs to be re-designed and re-instantiated as the
domain knowledge evolves.
%
%
%
%
In summary, \textit{prior-art control-plane generator SMM heavily relies on
domain knowledge which not only incurs significant manual effort but also
requires repeating the effort in reacting to changes and complications
in domain knowledge.}

\section{Synthesizing Control-plane Traffic without Domain Knowledge}
\label{sec:design}

To design a traffic generator that does not rely on domain-specific knowledge,
we leverage generative ML models which can be directly trained end-to-end on
raw traces, without needing domain knowledge such as 3GPP protocols.
We begin by exploring prior-art ML-based methods for generating
network time series, which have been shown to generate Internet data traffic
with high fidelity, to understand their limitations in generating cellular
network control-plane traffic traces.
We then present the design of \name, which addresses the limitations of
existing ML-based approaches.

\subsection{Dataset Overview}
\label{subsec:dataset_overview}

To enable our study of control-plane traffic generation, we collected a
cellular control-plane traffic dataset in collaboration with a major carrier in
the US.
The dataset is collected on the operator's LTE network by
randomly sampling from UEs across the entire U.S. over a span of 8 days in
2022.
In total, the dataset contains 73,153,370 control events, from 430,939 UEs
belonging to three device types: phones (278,389), connected cars (113,182),
and tablets (39,368).

\subsection{Prior-art ML-based Approach and Limitations}

Encouraged by the effectiveness of Generative Adversarial
Networks (GANs) applied to generating time sequences such as
NLP~\cite{chen2018adversarial}, systems researchers have applied GANs to
develop generators for networked
traffic~\cite{cheng2019pac,ring2019flow,doppelgan:imc2020,netshare:sigcomm2022,hui2022knowledge},
webpage views~\cite{doppelgan:imc2020}, and cluster job
traces~\cite{doppelgan:imc2020} with encouraging results.
Compared with classic simulation-based or statistical-based approaches, the
GAN-based approach requires minimal domain knowledge and thus has the potential
to support diverse types of network time series with little to no
customization.

DoppelGANger~\cite{doppelgan:imc2020} and NetShare~\cite{netshare:sigcomm2022} are
state-of-the-art GAN-based frameworks designed for network traffic generation.
Their key ingredient is to generate the metadata per stream (\eg 5-tuples of a TCP connection)
and the stream of samples (\eg IP headers)
conditioned on the metadata
separately, using an MLP-based generator 
and an LSTM-based generator, respectively.
Such a decoupled two-model architecture demonstrates superior fidelity over
several previous GAN-based traffic generators
(\eg~\cite{yoon2019time,esteban2017real}) in generating Internet data
traffic such as webpage access and cluster usage traces.
NetShare builds upon DoppelGANger and specializes in generating IP header and
flow traces. By incorporating domain knowledge
such as domain-specific encoding schemes (\eg bitwise
encoding~\cite{xu2019modeling} for IP addresses and
IP2Vec~\cite{ring2017ip2vec} for ports),
NetShare is shown to generate such traces with improved fidelity.
%
%
%

\subsubsection{Limitations of GAN-based Approach.}
\label{subsubsec:limitnetshare}

However, our evaluation reveals several limitations of prior-art
GAN-based approaches in
synthesizing cellular network control-plane traffic datasets.

\noindent
{\bf Adapting NetShare to control traffic.}
We adapt NetShare to synthesize cellular control-plane traffic as follows.
The original NetShare uses an MLP-based generator and LSTM-based generator to respectively generate a
``metadata'' and a ``time series'' that together form a stream,
where metadata refers
to fields that are universal to the entire stream (\eg the 5-tuple).
%
The equivalent metadata in cellular network control-plane traffic is the UE ID.
However, since a UE ID is a hashed string without semantic meaning, it is not
meaningful to generate it with a model or evaluate its fidelity. Instead, we
use a random string generator to generate UE IDs separately.

Therefore, we discard the metadata generator, and only use NetShare's
LSTM-based generator to generate time series where each sample in the time
series contains three fields: event type, interarrival time, and a ``stop
flag'' that indicates whether the current sample is the last in the stream.
\footnote{
NetShare could potentially be enhanced by replacing the LSTM model in its GAN
generator with a transformer model. However, as we will demonstrate with \name,
a transformer model can achieve sufficient performance with supervised
learning, eliminating the need for GAN-based training.
}
Other model hyperparameters of NetShare are kept unchanged.
For comparison, we present the dataset generated by both NetShare
and \name, with a detailed description of \name's design deferred to
\autoref{sec:design}.
%
The model is trained following the methodologies that will be described in
\autoref{subsec:evaluation_setup},
and the inference setup is deferred to \autoref{subsec:evaluation_setup},
\autoref{subsec:overall_results} and \autoref{subsec:adapting_to_data_drifts}.


\boldbox{Limitation 1 (L1): State-of-the-art ML-based traffic generators do not
achieve stateful semantic correctness.}

\begin{table}[tp]
\caption{Semantic violations in control-plane traffic synthesized by NetShare.}
\label{tab:violation}
\resizebox{0.8\columnwidth}{!}{
\begin{tabular}{lr}
\toprule
\multicolumn{1}{l|}{Perc. event violations} & 2.61\% \\
\multicolumn{1}{l|}{Perc. streams w/ at least one violating event} & 22.10\% \\ \midrule
\multicolumn{2}{c}{State and event of top 3 violation} \\ \hline
\multicolumn{1}{l|}{\texttt{S1\_REL\_S}, \texttt{S1\_CONN\_REL}} & 1.16\% \\
\multicolumn{1}{l|}{\texttt{S1\_REL\_S}, \texttt{HO}} & 0.76\% \\
\multicolumn{1}{l|}{\texttt{CONNECTED}, \texttt{SRV\_REQ}} & 0.41\% \\ \bottomrule
\end{tabular}
}
\end{table}

Many network datasets exhibit stateful
semantics~\cite{fonseca2007x,kim2015kinetic,pontarelli2019flowblaze,meng2023modeling},
because they are usually produced by end hosts or middleboxes that operate
according to some network protocols, upper-layer application rules, or
middlebox policies.
In the case of cellular control-plane traffic, the stream of control events
produced by each UE is the outcome of a series of state transitions on the UE
state machine derived from the 3GPP standard (Fig.~\ref{fig:lte_5g_sm}).

We evaluate the semantic correctness of NetShare-generated trace by feeding
each stream into the 3GPP state machine, and count how many events violate the state
transitions.
~\autoref{tab:violation} shows that 2.614\% of the events in
the dataset synthesized by NetShare violate the 3GPP-defined state
transitions.
Such a degree of event violations translates to 22.10\% of the streams
generated containing at least one semantic-violating events, rendering these
streams potentially useless for downstream applications.
A detailed analysis shows the reason for such a high degree of stateful semantics violations
is that NetShare fails to capture the fact that
\texttt{S1\_CONN\_REL} and \texttt{HO} events should not occur when the UE is
in \texttt{S1\_REL\_S} state (Fig.~\ref{fig:lte_sm}).


\boldbox{Limitation 2 (L2): State-of-the-art ML-based traffic generators do not
synthesize high-fidelity traces in terms of sojourn times.}

\begin{figure}[tp]
    \centering
    \includegraphics[width=0.7\columnwidth]{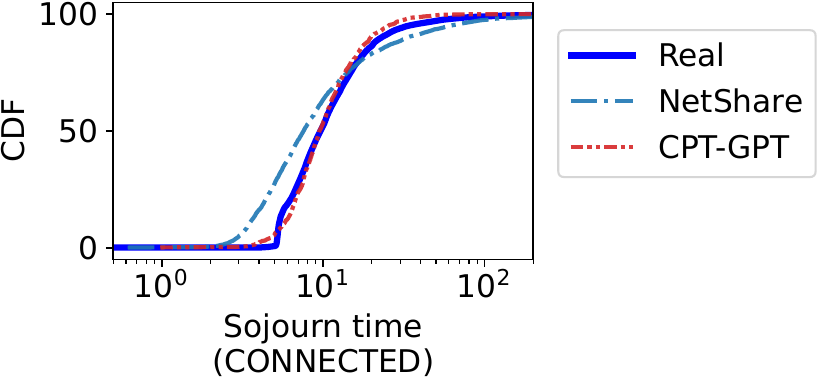}
    \caption{
        Distributions of the average sojourn time in the \texttt{CONNECTED} of
        each UE, comparing real and synthesized traces, for phone UEs.
    }
    \label{fig:cdfs}
\end{figure}

\autoref{fig:cdfs} compares the CDF of the average sojourn times per UE in
the \texttt{CONNECTED} state for each synthesized dataset as well as the real
dataset, for phone UEs.
We see that NetShare does not perform as well in
generating high fidelity sojourn times.
%
Specifically, while the majority of streams in the real dataset have an averaged
\texttt{CONNECTED} state sojourn time ranging from 5 to 50 seconds, NetShare
frequently generates streams with sojourn times spanning 2 to 100 seconds
with a non-negligible probability.
This results in a significant difference in the maximum y-distance between the
CDFs of the synthesized and real dataset, at 27.9\% and 6.4\% for NetShare and
\name, respectively.
We refer readers to \autoref{subsec:overall_results} for the results for the
other two types of UEs.

\boldbox{Limitation 3 (L3): State-of-the-art ML-based traffic generators cannot
efficiently adapt to different workloads via transfer learning.}

Just like Internet data traffic, the control-plane traffic of cellular
networks generated by UEs exhibits variations over time, which can be attributed to 
diurnal, weekly, or seasonal fluctuations in UE behavior,
as well as
network changes resulting from the continuous cellular network evolution
or incremental deployment of a given generation of cellular networks.
Therefore, the traffic generation framework
may need to re-train a new ML model that captures the resulting variations in
control-plane traffic, \eg once every month.
To reduce the model training cost for such drifting workloads,
instead of training a new ensemble of hourly models from scratch, a
potentially efficient approach is to
%
{train a base model for a specific hour-of-day from scratch, and then adapt it
using transfer learning to generate a model for each remaining hour of the
day~\cite{weiss2016survey,radford2018improving,bhardwaj2022ekya}.
}

\begin{table}[tp]
\caption{Time needed to train NetShare from scratch, or perform transfer
learning on an existing model to adapt to different workloads. Training stops
when fidelity metrics show diminishing returns. Measurement is done on an
NVIDIA A100 GPU.}
\label{tab:training_time_netshare}
\resizebox{\columnwidth}{!}{
\begin{tabular}{l|c}
\toprule
6-hour model from scratch & 108.36 mins \\ \midrule
1-hour model from scratch & 43.08 mins \\
1-hour model from finetuning from another hour & 30.41 mins \\
6 1-hour models total from transfer learning & 195.12 mins \\ \bottomrule
\end{tabular}
}
\end{table}

However, we found that state-of-the-art GAN-based frameworks are inefficient in
performing such transfer learning.
We measure NetShare's training time on trace collected over 6 hours for the
phone device type,
and compare it with training NetShare just on the first hour of the trace and
then re-trained to adapt to the five subsequent hours recursively (we refer
readers to \autoref{subsec:adapting_to_data_drifts} for the detailed
methodology).
\autoref{tab:training_time_netshare} compares the training time for the two
approaches. We observe that it takes 108.36 minutes to train a single NetShare
model covering 6 hours of traces from scratch.
In contrast, the time needed to train an ensemble of specialized models via
transfer learning is 195.12 minutes, almost twice as long. This indicates that
the GAN/LSTM design cannot leverage transfer learning to adapt to shifting
workloads effectively, which has been shown in prior
works~\cite{wang2018transferring,noguchi2019image,wang2020minegan}.

\boldbox{Limitation 4 (L4): State-of-the-art ML-based traffic generators use
LSTMs which are insufficient in learning long-term dependencies in the traces.}

DoppelGANger and NetShare use LSTMs to 
capture dependencies in sequential data. However, using off-the-shelf LSTMs to
synthesize control plane traffic, \ie one step at a time,  
suffers from an inherent limitation to forget states over long sequences, as
shown by authors of DoppelGANger and
NetShare~\cite{doppelgan:imc2020,netshare:sigcomm2022}.
%
To overcome this problem, DoppelGANger and NetShare employ batch generation,
\ie generating multiple samples in each LSTM step, thereby reducing the number
of passes to produce a stream of a certain length and hence the amount
of forgotten states.
%
%
However, since batch generation produces multiple samples each step, it
sacrifices intra-batch dependencies, which could compromise fidelity, as
each cellular control event depends on the event immediately preceding it.
This potentially contributed to the low fidelity measures in L1 -- L3.
  
\boldbox{Limitation 5 (L5): State-of-the-art ML-based traffic generators use
GANs which suffer from mode collapse.}

GAN training is known to suffer from mode
collapse~\cite{kodali2017convergence,thanh2020catastrophic,srivastava2017veegan},
where the generator learns to generate only one or few plausible outputs,
rather than a wide variety of outputs as in the training dataset.
Rather than using general solutions for mode collapse, DoppelGANger and
NetShare alleviate mode collapse with a specialized normalization scheme.
Specifically, each stream is normalized individually, using the min and max
values within the stream, rather than a global min/max value.
This shows off-the-shelf GAN-based traffic generators are susceptible to mode
collapse, and require
tailored, traffic type- and model-dependent enhancements to mitigate the
issue.

\subsection{Why Transformers?}

A transformer~\cite{vaswani2017attention} is a deep learning architecture built around
the ``attention'' mechanism. Input text is represented as a sequence of tokens,
and at each layer, an attention score is computed between every pair of tokens
through a multi-head attention mechanism.
The attention mechanism facilitates the learning of dependencies
between distant tokens, in contrast to RNNs or LSTMs which process a
stream of tokens sequentially.  Moreover, such design allows for the
parallel processing of the tokens which are highly efficient on
today's hardware accelerators such as GPUs.
%
%

Transformers have the potential to address the aforementioned limitations
(L1-L5) of existing ML-based traffic generators.
{
(1) Transformers have been shown to perform well for a variety of generative
tasks, such as generating code that conforms to a rigid grammar
(L1)~\cite{katinskaia2021assessing}, as well as capturing complex dependencies
among multiple fields in tokens in a stream (L2) for multi-modality tasks such
as music generation~\cite{huang2018music}.
(2) Furthermore, since their attention mechanism allows the input sequence to
be processed in parallel, they have the potential to reduce the training time
on today's hardware, compared to LSTMs which process the input sequentially
(L3)~\cite{vaswani2017attention}.
(3) Transformer is known to have superior performance than LSTMs on learning
long-term dependencies (L4)~\cite{karita2019comparative}, as its attention
mechanism models the dependencies without regard to their distance in the
sequence.
(4) Finally, they are typically trained in a supervised fashion and thus do not
suffer from mode collapse like GANs (L5)~\cite{kodali2017convergence}.
}



\subsection{\name Design}
\label{subsec:design}

\boldbox{Design 1: We employ a multi-modal tokenization scheme tailored for
cellular network control traffic.}

%
The powerful attention mechanism at the core of transformers was originally
designed to capture pair-wise dependencies of {\em finite-sized,
single-modality} tokens in NLP tasks that process streams of words from a
finite set of vocabulary, \eg the state-of-the-art Llama-3~\cite{llama3}
employs a vocabulary size of 32,000.
%
The finite-sized vocabulary of words allows them to be effectively encoded into one-hot tokens.
In contrast, in network traffic, streams of samples, \eg UE control events in
cellular networks, contain multiple modalities (\eg continuous interarrival time and
categorical event types in control-plane traffic),
and furthermore, the continuous field has a significantly larger input space than
categorical fields.
Together the mixed types of fields in network traffic
result in a significantly larger input and output space than NLP tasks,
rendering language-style one-hot tokenization inapplicable to network traffic.

Similarly, vision transformers designed for CV
tasks~\cite{dosovitskiy2010image} convert image patches to floating point
matrices and then encode them into tokens, which also have a single modality
although a large token space. In comparison, network traffic has multiple
modalities of different types, either continuous or numerical, that need to be
treated differently, and thus the tokenization scheme in vision transformers
cannot be easily applied in network traffic either.
%

While one may concatenate multiple fields in each sample of network
traffic stream and treat them as a single field, it is not clear
whether language or vision transformers can learn the distribution of
each individual field or capture the correlation across the fields.

\begin{figure}[tp]
    \includegraphics[width=0.6\columnwidth]{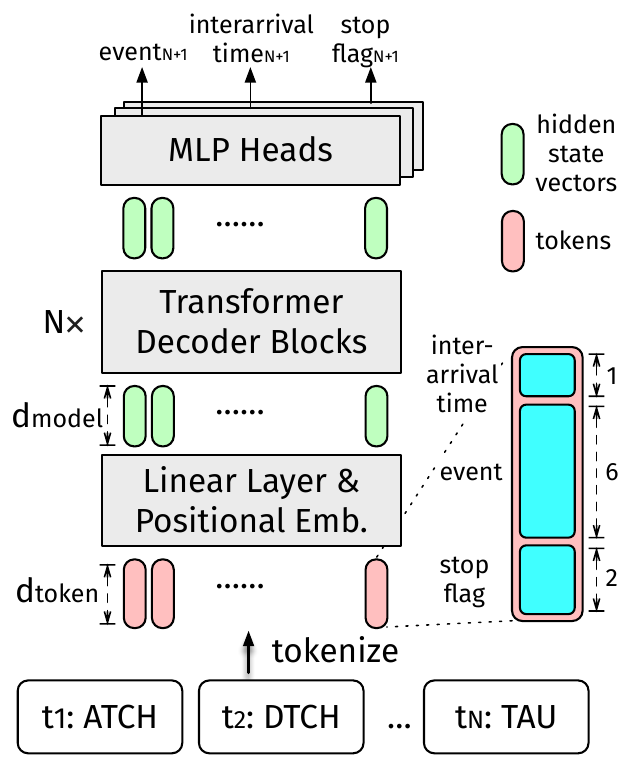}
    \caption{\name architecture and tokenization scheme.}
    \label{fig:model}
\end{figure}

\parb{Our solution}
To integrate multi-modality data with the transformer architecture, we designed
a specialized tokenization strategy where each token is the concatenation of
multiple sub-tokens, each representing a specific modality, as illustrated in
\autoref{fig:model}.
For categorical fields, such as event type, we convert them into sub-tokens
using one-hot encoding.
Numerical fields, such as interarrival time, are applied log scaling (\ie
$x'=log(x+1)$) and then linear scaled to a range of 0 -- 1, 
where 0 and 1 correspond to the minimum and maximum interarrival times across the
dataset\footnote{
The rationale behind log scaling is that interarrival time spans several orders
of magnitude, where the majority of the data is concentrated in smaller ranges,
as shown in \autoref{fig:interarrival_cdf_phone} in
\autoref{sec:additional_dataset_statistics}.
Log scaling expands the range of smaller values and compresses the range of
larger ones, making the data more uniformly distributed.
}.
Additionally, generating streams of varying lengths requires an additional
``stop flag'' sub-token, marking whether the token is the last in a stream. The
stop flag is a categorical field taking values of zero or one,
similarly as in NetShare~\cite{netshare:sigcomm2022}.

Correspondingly, as shown in \autoref{fig:model}, the ``embedding'' layer in
classic NLP transformers is replaced with a linear layer that maps a
$d_{token}$-dimension space to a $d_{model}$-dimension space, where $d_{token}$ is the
dimension of the tokens (9 in our case, concatenating the three sub-tokens
1+6+2), and $d_{model}$ is a model hyperparameter representing the attention
hidden size.
Furthermore, to produce multi-modality outputs, \name employs three multi-layer
perception (MLP) heads following the final attention block,
each making predictions for a particular field. 

\boldbox{Design 2: We design the model to output distribution parameters for
numerical fields to improve generation stochasticity.}

Model architecture design for producing multi-modality output is fairly
standard.
For instance, in NetShare, each modality is assigned its own MLP head.
For fields of a categorical modality, such as event type or the ``stop flag'',
the MLP head uses a SoftMax in the last layer to output a vector representing
the probability distribution over possible values.
During inference, the output is selected either by sampling from the
probability distribution, as commonly done in language
models~\cite{radford2019language}, or by simply choosing the element with the
highest possibility, as in NetShare.

However, the interarrival time field in cellular control-plane traffic is of
numerical modality with continuous values.
Previous designs, such as NetShare, output a single numerical value from the
corresponding MLP head.
But as we will show in the ablation study
(\autoref{subsec:ablation_and_sensitivity}), such a design results in a lack of
generation stochasticity, \ie the model tends to generate the mean value instead of
producing a diverse range of values, failing to address C3 \& C4.
Overcoming this limitation requires designing alternative mechanisms to ensure 
generation stochasticity for such numerical fields.




\parb{Our solution}
  We enhance the MLP head for the numerical field in \name
(\ie interarrival time) by configuring it to output the
parameters of a probability distribution,
rather than a single numerical value.
We empirically chose normal distribution and make the model output both a mean
and a standard deviation.
During training, the Gaussian negative log-likelihood (NLL) loss function is
used for numerical fields, while categorical fields continue to use
cross-entropy loss,
and the training minimizes the weighted sum of these losses across fields.
During inference, values for numerical fields are randomly sampled from the
predicted probability distributions, similarly as in the approach used for
categorical fields.

\vfill\eject     
\boldbox{Design 3: We employ transfer learning to efficiently generate models
for network traffic shifts.}

As mentioned in \autoref{subsec:overview_challenges} (C5), the ML model needs to
be updated to reflect time-varying characteristics in the dataset, stemming from short-term UE
behavior fluctuations (\eg diurnal and weekly), as well as longer-term changes due
to evolving cellular network deployments or technologies.
Therefore, the traffic generation framework needs to
periodically retrain the model on the latest
collected network traces,
which potentially incurs significant
training costs.

\parb{Our solution}
To reduce the training time needed to train new models as the dataset
characteristics drift through time, we employ a transfer learning strategy
where the transformer model is first trained on a specific hour, and then
we perform transfer learning on it to generate models recursively for each
subsequent hour.
Such a transfer-learning scheme would provide significant time savings, as we
will show that training based on a pretrained model from another hour is much
faster than training a model from scratch.

\subsection{Architecture Overview}
\label{subsec:architecture_overview}

\begin{figure}[tp]
    \includegraphics[width=0.75\columnwidth]{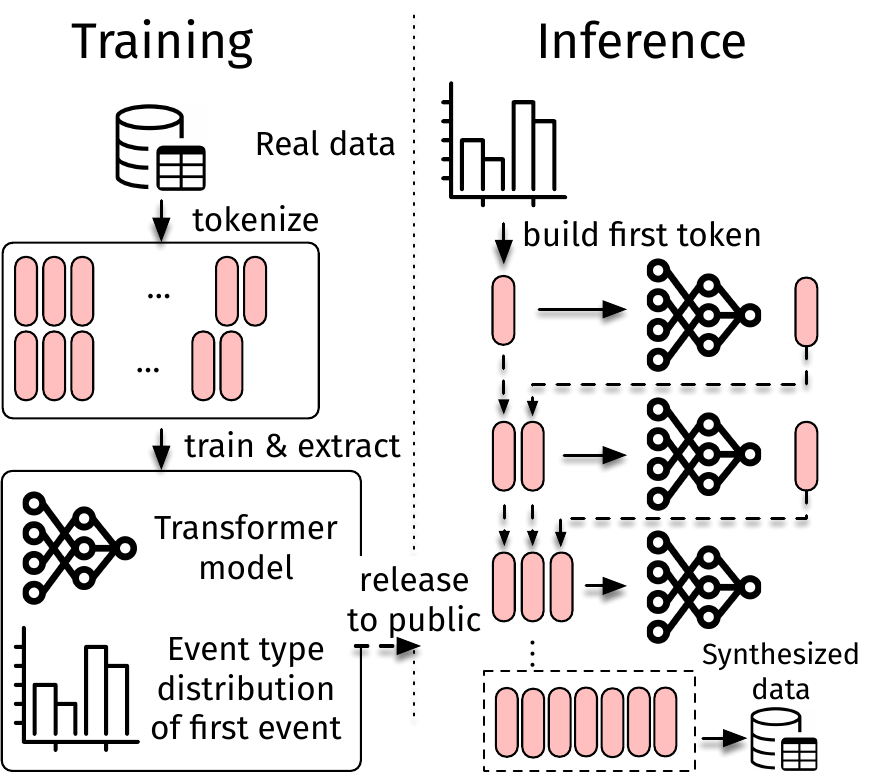}
    \caption{Operational architecture overview.}
    \label{fig:architecture_overview}
\end{figure}

Combining the design insights above, the overall architecture and workflow of \name are shown in
\autoref{fig:architecture_overview}.

\parb{Training} The traffic generation framework developer, \eg a
cellular operator, collects a control-plane traffic dataset and
applies \name's tokenization scheme to tokenize it, and subsequently
uses it to train the transformer model.  Additionally, the developer
extracts the event type distribution of the initial events across all
streams in the dataset, which will be used to bootstrap the \name
inference as explained in the next paragraph.
The trained model weights, along with the initial-event-type distribution, will be
packaged together and released to the public.

\parb{Inference}
To synthesize a control-plane traffic trace for a UE population of size $N$,
a \name user invokes the transformer-decoder-based \name $N$ times.

In each inference,
\name first randomly samples an event type from the released initial-event-type distribution.
It then constructs 
a token with the event type
and with the remaining two fields, \ie the interarrival time and the stop
flag, set to zero. This is consistent with model training where the first
token always has an interarrival time and stop flag of zero (streams of length
1, \ie with the first token having a stop flag of 1, are excluded from the
training).
Finally, \name uses the token as the prompt to bootstrap the inference; it runs
\name inference recursively, predicting the (K+1)-th token in the stream based
on previous K tokens, until encountering a token containing a stop flag of 1.

\section{Evaluation}

\subsection{Evaluation Setup}
\label{subsec:evaluation_setup}

\begin{table*}[tp]
\caption{Percentage of events and streams that violate the stateful semantics
defined by 3GPP standards. The results for SMM are omitted as it does not
generate any state violations.}
\label{tab:violations}
\resizebox{0.75\textwidth}{!}{
\begin{tabular}{l|cc|cc|cc}
\toprule
 & \multicolumn{2}{c|}{Phone} & \multicolumn{2}{c|}{Connected Car} & \multicolumn{2}{c}{Tablets} \\
 & NetShare & \multicolumn{1}{l|}{\name} & NetShare & \name & NetShare & \name \\ \midrule
Event violations (\%) & 2.614\% & 0.004\% & 3.915\% & 0.034\% & 3.572\% & 0.079\% \\
Streams w/ at least one violating event (\%) & 22.1\% & 0.2\% & 11.5\% & 0.4\% & 16.9\% & 1.5\% \\ \bottomrule
\end{tabular}
}
\end{table*}

\parbi{Baselines}
We compare \name with both the traditional modeling-based 
and the state-of-the-art GAN-based approaches, \ie SMM~\cite{meng2023modeling}
and NetShare~\cite{doppelgan:imc2020}.
Considering SMM requires more than 20K Semi-Markov models and 200K CDFs to cover
all 24 hours of the day and the three device types, which is extremely
cumbersome to derive and use, we also consider a variation of SMM (denoted as
SMM-1), which generates a single Semi-Markov model per device type, and the
original SMM model is referred to as SMM-20k.
We already described in detail how NetShare was adapted to synthesize
control-plane traffic in \S\ref{subsubsec:limitnetshare}.

\parbi{Training}
We use a dataset collected over seven days in June 2022 as the training set,
and use another dataset collected over one day in August 2022 for testing.
%
%
The trace for the same UE across different days is treated as different UEs,
and the 24-hour-long traces are divided into 24 traces of one hour in length
each.
Furthermore, both NetShare and \name generate sequences up to a preconfigured
maximum length. To enable a fair comparison between them, we configure and
train their models to synthesize streams with a maximum length of 500,
disregarding those exceeding this threshold (only 0.07\% of the
1-hour flows exceed a length of 500).
%
%
We tuned the hyperparameters for \name to identify the largest model that does
not overfit on the testing set with the ``phone'' device type. The tuning
process led to models with 2 attention blocks, each with embedding dimension
of 128 and MLP hidden size of 1024, for a total of 725K parameters and a weight
size of 2.9 MB.

We initially train a NetShare and a \name model from scratch on
the subset of the dataset recorded by UEs with the ``phone'' device type.
We then continue training these models using transfer learning (discussed in
\autoref{subsec:design}) to adapt them to the other two device types, connected
cars and tablets.

\parbi{Inference}
The fidelity evaluation involves synthesizing 1000 UE streams for every 
traffic generator.
%

\parbi{Time measurements}
To ensure consistent runtime measurements across models, all experiments are
run on servers equipped with an NVIDIA A100 GPU, an AMD EPYC 7543 CPU, and 256
GB of memory.

\subsection{Overall Results}
\label{subsec:overall_results}

\subsubsection{Stateful Semantic Correctness.}

\begin{table*}[tp]
\caption{{Maximum y-distance between the CDFs of the real and synthesized dataset.}}
\label{tab:max_y_dist}
\resizebox{1.0\textwidth}{!}{
\begin{tabular}{cc|cccc|cccc|cccc}
\toprule
\multicolumn{1}{l}{} & \multicolumn{1}{l|}{} & \multicolumn{4}{c|}{Phone} & \multicolumn{4}{c|}{Connected Car} & \multicolumn{4}{c}{Tablet} \\
\multicolumn{1}{l}{} & \multicolumn{1}{l|}{} & SMM-1 & \multicolumn{1}{l}{SMM-20k} & NetShare & Ours & SMM-1 & \multicolumn{1}{l}{SMM-20k} & NetShare & Ours & SMM-1 & \multicolumn{1}{l}{SMM-20k} & NetShare & Ours \\ \midrule
\multicolumn{1}{c|}{\multirow{2}{*}{Sojourn time}} & \texttt{CONNECTED} & 40.1\% & 14.8\% & 27.9\% & 6.4\% & 45.1\% & 16.8\% & 61.7\% & 26.4\% & 44.0\% & 17.6\% & 53.6\% & 11.3\% \\
\multicolumn{1}{c|}{} & \texttt{IDLE} & 37.6\% & 9.6\% & 12.0\% & 12.0\% & 46.8\% & 14.8\% & 16.2\% & 33.3\% & 35.5\% & 15.4\% & 25.7\% & 11.5\% \\ \midrule
\multicolumn{1}{c|}{\multirow{3}{*}{Flow length}} & All events & 44.2\% & 1.9\% & 1.6\% & 3.8\% & 54.7\% & 9.6\% & 1.4\% & 4.5\% & 60.2\% & 18.7\% & 3.8\% & 3.6\% \\
\multicolumn{1}{c|}{} & \texttt{SRV\_REQ} & 41.9\% & 3.7\% & 2.4\% & 4.3\% & 55.4\% & 9.7\% & 4.0\% & 5.9\% & 56.5\% & 13.1\% & 4.4\% & 5.0\% \\
\multicolumn{1}{c|}{} & \texttt{S1\_CONN\_REL} & 43.5\% & 1.7\% & 1.5\% & 4.0\% & 56.0\% & 7.1\% & 3.5\% & 5.0\% & 60.0\% & 18.3\% & 3.4\% & 3.5\% \\ \bottomrule
\end{tabular}
}
\end{table*}

We first quantify the semantic correctness of the synthesized dataset as follows. For
each synthesized stream, we sequentially replay the events against the UE state
machine (\autoref{fig:lte_sm}). On encountering a state-violating event, a
counter is incremented and the state machine stays in the same state. To
bootstrap the initial state of the state machine, we employ a heuristic that
looks for the first \texttt{ATCH}, \texttt{DTCH}, \texttt{SRV\_REQ}, or
\texttt{HO} event, as these events lead to a deterministic destination UE state
regardless of the source state and thus can be used to determine the subsequent
states. Events preceding the state machine bootstrapping are excluded from the
semantic correctness calculation.

{\autoref{tab:violations} shows the percentage of state-violating events
as well as the percentage of streams that contain at least one state-violating
event.}
{Since SMM-1 and SMM-20k have state machines built-in, they naturally do
not generate any state-violating events and thus their results are not shown.}
In terms of the percentage of events that violate the stateful semantics, \name
consistently results in minimal violations, at 0.004\%, 0.034\%, and 0.079\%
for the three device types, respectively.
This is two order-of-magnitude lower than that of NetShare, which results in
2.614\%, 3.915\%, and 3.572\% events triggering violations.

The percentage of stream violations is naturally higher than event violations
for both models, as a stream is deemed to violate the semantics if it contains
at least one violating event.
However, the stark contrast between NetShare and \name remains. NetShare
generates 22.1\%, 11.5\%, and 16.9\% streams with violations for the three
device types respectively, while \name results in much lower violations of
0.2\%, 0.4\%, and 1.5\%.
This shows the transformer-based \name is highly effective in comprehending and
generating control-plane traffic with stateful semantics autonomously, without
reliance on domain knowledge as in SMM-1 and SMM-20k, a strength that NetShare
fails to achieve.

\subsubsection{Distribution Metrics.}

Next, we compare the fidelity of the synthesized datasets of different
approaches in terms of the three distribution-related metrics shown in
\autoref{tab:metrics}, namely, sojourn time distribution, event breakdown,
and flow length distribution.

{
\parbi{Sojourn time}
\autoref{tab:max_y_dist} shows the maximum y-distances of the sojourn time CDFs
for each synthesized dataset and the real dataset, for both the
\texttt{CONNECTED} and \texttt{IDLE} states.
First, compared with SMM-20k, \name achieves comparable fidelity
out of the 6 scenarios (2 states $\times$ 3 device types).
Specifically, \name outperforms in 3 scenarios, for both states with tablet
UEs (11.3\% v.s. 17.6\% for \texttt{CONNECTED}, 11.5\% v.s. 15.4\$ for
\texttt{IDLE}) as well as in the \texttt{CONNECTED} state with phone UEs
(6.4\% v.s. 14.8\%), while achieving lower accuracy in the remaining 3
scenarios.
Second, compared to SMM-1 and NetShare, \name achieves superior
fidelity in the \texttt{CONNECTED} state sojourn time across all 3 device
types, with maximum y-distances of 6.4\% / 26.4\% / 11.3\% respectively,
significantly lower than SMM-1's 40.1\% / 45.1\% / 44.0\%, and NetShare's
27.9\% / 61.7\% / 53.6\%.
%
For \texttt{IDLE} state sojourn time, \name's  consistently outperforms SMM-1
across all device types and shows comparable fidelity to NetShare (\ie better
for tablets, equal for phones, and worse for connected cars).

\begin{figure*}[tp]
    \centering
    \includegraphics[width=1.0\textwidth]{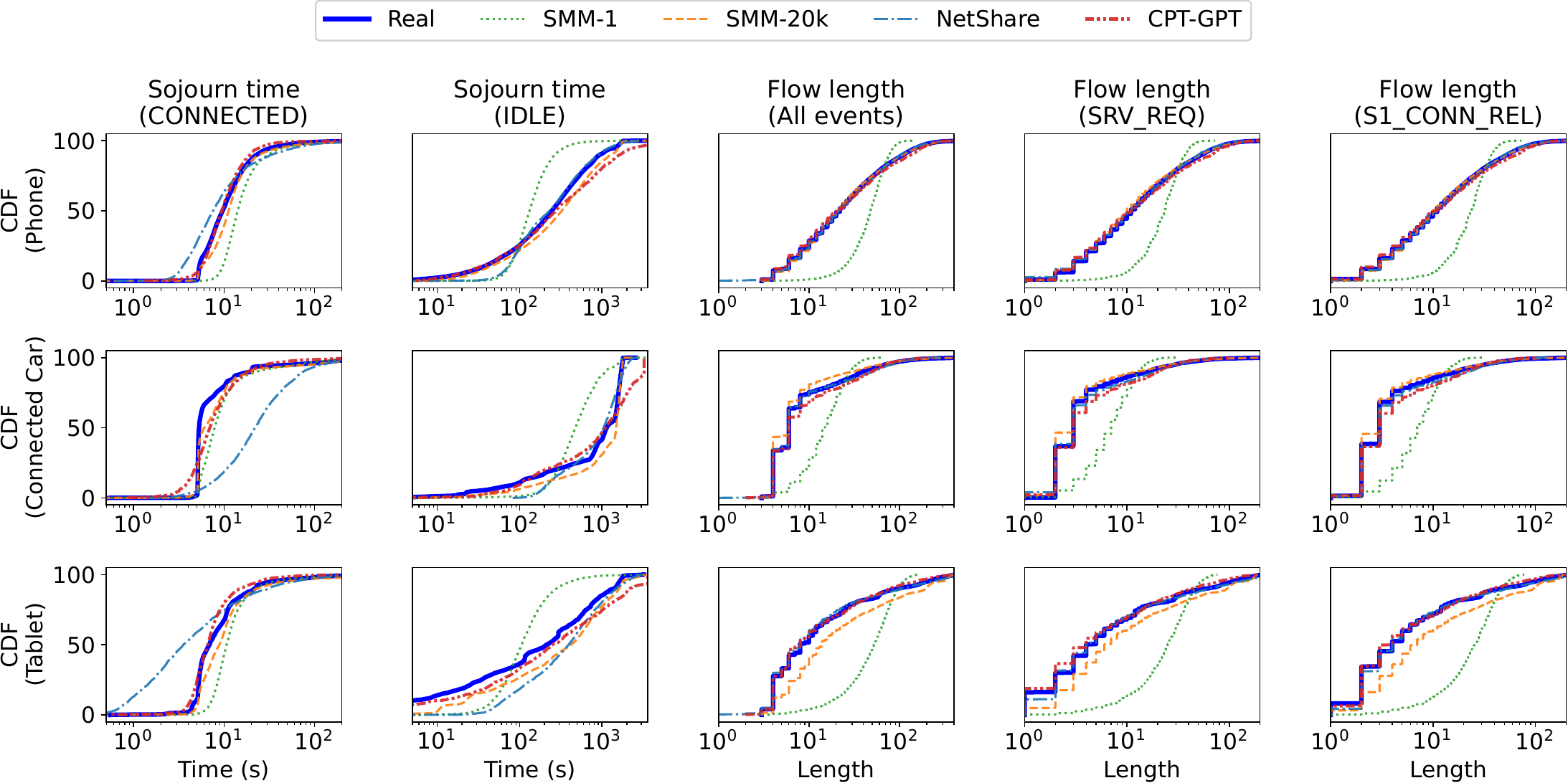}
    \caption{Distributions of fidelity metrics for different types of UEs.}
    \label{fig:cdfs_eval_sec}
\end{figure*}

In \autoref{fig:cdfs_eval_sec}, we show the actual CDFs for different types of UEs.
Examining the sojourn time distributions (left two columns), we observe that:
%
(1) Generally, \name and SMM-20k are the two most accurate generators,
with distributions most closely mirroring that of the real dataset 
across all device types.
(2) SMM-1 is the least accurate generator. For the \texttt{CONNECTED} state,
it tends to generate longer-than-realistic sojourn times for
phones and connected cars, but shorter-than-realistic times for tablets.
For the \texttt{IDLE} state, it tends to generate too many
sojourn times that are around 100 seconds for phones and tablets
and too many sojourn times of around 200-300 seconds for connected cars.
(3) While NetShare generates \texttt{IDLE} sojourn times as realistic as \name,
it generates much less accurate \texttt{CONNECTED} state
sojourn times, especially for
connected cars and tablets, where it tends to generate too long and too
short sojourn times, respectively.

In summary, \name achieves superior sojourn time fidelity compared to NetShare
and SMM-1 in most scenarios --
{
on average, the average max y-distance
over the two 3GPP states and three device types
is reduced by 24.7\% and 16.0\%, respectively.
More importantly, \name achieves comparable accuracy (within 2.0\% max y-distance)
as SMM-20k.
}

\parbi{Flow length}
\autoref{fig:cdfs_eval_sec} (middle column) shows the distribution of the
number of events per stream. 
We refer readers to \autoref{tab:max_y_dist} which shows the maximum
y-distances between the CDFs of real and synthesized dataset, with the values
presented in percentages, where a smaller difference implies better accuracy in
replicating the distribution of the real dataset.
In addition to the flow length across all event types, we specifically
highlight the length for \texttt{SRV\_REQ} and \texttt{S1\_CONN\_REL} events
(\ie the number of respective events in each stream), the two dominant event
types, in the right two columns in \autoref{fig:cdfs_eval_sec}.

\begin{table*}[tp]
\caption{{Breakdown of event types of real dataset, and of each
synthesized dataset shown as difference compared to the real dataset. A lower
difference is more accurate.}}
\label{tab:macropscopic}
\resizebox{1.0\textwidth}{!}{
\begin{tabular}{c|ccccc|ccccc|ccccc}
\toprule
 & \multicolumn{5}{c|}{Phones} & \multicolumn{5}{c|}{Connected Car} & \multicolumn{5}{c}{Tablets} \\
 & \multicolumn{1}{c|}{Real} & SMM-1 & \multicolumn{1}{l}{SMM-20k} & NetShare & Ours & \multicolumn{1}{c|}{Real} & SMM-1 & \multicolumn{1}{l}{SMM-20k} & NetShare & Ours & \multicolumn{1}{c|}{Real} & SMM-1 & \multicolumn{1}{l}{SMM-20k} & NetShare & Ours \\ \midrule
\texttt{ATCH} & \multicolumn{1}{c|}{0.12\%} & -0.01\% & 0.02\% & -0.09\% & -0.01\% & \multicolumn{1}{c|}{1.00\%} & -0.04\% & 0.14\% & -0.61\% & -0.12\% & \multicolumn{1}{c|}{1.13\%} & -0.50\% & -0.37\% & -0.70\% & 0.06\% \\
\texttt{DTCH} & \multicolumn{1}{c|}{0.11\%} & 0.01\% & 0.04\% & -0.05\% & -0.02\% & \multicolumn{1}{c|}{0.97\%} & 0.02\% & 0.21\% & -0.53\% & -0.18\% & \multicolumn{1}{c|}{1.08\%} & -0.45\% & -0.29\% & -0.63\% & -0.08\% \\
\texttt{SRV\_REQ} & \multicolumn{1}{c|}{47.06\%} & 0.99\% & 0.75\% & 0.28\% & 0.66\% & \multicolumn{1}{c|}{39.75\%} & 6.11\% & 5.67\% & 1.38\% & 2.15\% & \multicolumn{1}{c|}{44.51\%} & 3.52\% & 3.03\% & 2.26\% & -3.62\% \\
\texttt{S1\_CONN\_REL} & \multicolumn{1}{c|}{48.25\%} & 0.69\% & 0.64\% & 0.43\% & 0.34\% & \multicolumn{1}{c|}{44.14\%} & 3.63\% & 3.30\% & 0.08\% & 0.96\% & \multicolumn{1}{c|}{47.70\%} & 1.22\% & 1.03\% & 1.49\% & 0.03\% \\
\texttt{HO} & \multicolumn{1}{c|}{2.88\%} & -1.13\% & -1.12\% & 0.42\% & -0.50\% & \multicolumn{1}{c|}{8.59\%} & -6.17\% & -5.92\% & 1.23\% & -1.70\% & \multicolumn{1}{c|}{2.61\%} & -1.50\% & -1.36\% & -1.41\% & 0.00\% \\
\texttt{TAU} & \multicolumn{1}{c|}{1.59\%} & -0.56\% & -0.35\% & -1.00\% & -0.47\% & \multicolumn{1}{c|}{5.55\%} & -3.55\% & -3.40\% & -1.55\% & -1.12\% & \multicolumn{1}{c|}{2.97\%} & -2.29\% & -2.05\% & -1.01\% & 3.61\% \\ \bottomrule
\end{tabular}
}
\end{table*}

For the flow length consisting of all events (see ``All events'' in
\autoref{tab:max_y_dist}),
we observe the following:
(1)
\name and NetShare are generally rank as the top two generators in terms of
flow length distribution fidelity, with maximum y-distances of 3.6\%--4.5\% and
1.4\%--3.8\% respectively.
Although NetShare shows superior fidelity than \name on 2 out of 3 device types
(phones and connected cars), the differences are marginal (2.2\% and 3.1\%)
when compared with the discrepancies observed with the other generators.
(2) 
SMM-1 produces highly inaccurate
flow length distribution, resulting in a maximum y-distance of 44.2\%,
54.7\%, and 60.2\% for the three device types, respectively,
showing that a single Semi-Markov model is insufficient in
accurately modeling the flow length distributions.
{
The reason for such low fidelity, as shown in \autoref{fig:cdfs_eval_sec}, is
that SMM-1 generates excessive flows of length between approximately
20 to 100 for phones and tablets and 10 to 20 connected cars.
(3)
SMM-20k, which utilizes over 20k Semi-Markov models, generates much more
accurate flow length distributions than SMM-1, with maximum y-distances of
1.9\%, 9.6\%, and 18.7\% respectively.
However, while SMM-20k achieves a high fidelity of 1.9\% that is comparable to
NetShare and CPT-GPT for phones, its accuracy for connected cars and tablets is
significantly lower, at 9.6\% and 18.7\%, respectively.
As shown in \autoref{fig:cdfs_eval_sec}, SMM-20k performs poorly on connected
cars and tablets as it tends to generate too short and too long flows for the
two device types, respectively.
%
%
%
%

For the flow length of the two dominating events \texttt{SRV\_REQ} and
\texttt{S1\_CONN\_REL},
\autoref{tab:max_y_dist}
shows that 
\name and NetShare
remain as the top two generators, both demonstrating significantly higher
fidelity than SMM-1, and for connected cars and tablets, also outperforming
SMM-20k.
For example, in the case of phones, NetShare produces maximum CDF y-distances
of 2.4\% and 1.5\% for the two event types respectively, and \name results in
4.3\% and 2.4\%.
While NetShare is slightly more accurate than \name, their difference is
marginal when compared SMM-1, which yields maximum y-distances of 41.9\% and
43.5\%;
as \autoref{fig:cdfs_eval_sec} shows, SMM-1 fails to learn the distribution of
flow lengths and tends to generate too many flows of length 100-200 for phones
and tablets, and 50-100 for connected cars.
%
When compared to SMM-20k, \name and NetShare again demonstrate comparable
accuracy for phones (2.4\%/4.3\% v.s. 3.7\%), and superior accuracy for
connected cars (4.0\%/5.9\% v.s. 9.7\%) and tablets (4.4\%/5.0\% v.s. 13.1\%)
as \autoref{fig:cdfs_eval_sec} shows, SMM-20k tends to generate flows that are
too short for connected cars and too long for tablets.

In summary, both \name and NetShare demonstrate high fidelity in flow length
modeling, significantly outperforming other generators.
While NetShare outperforms \name in more scenarios, the difference is marginal.




\begin{figure*}[tp]
    \centering
    \includegraphics[width=\textwidth]{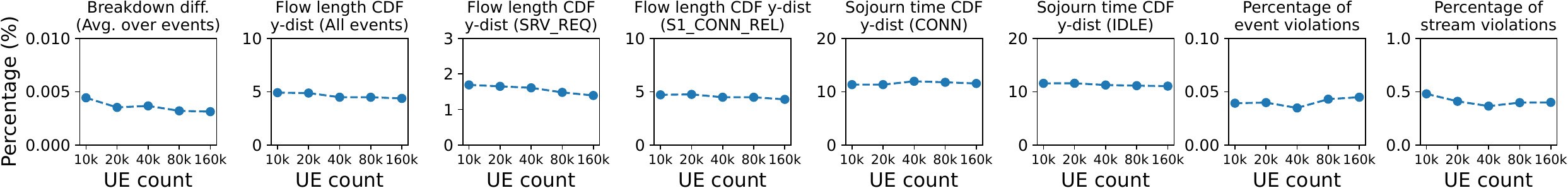}
    \caption{Fidelity of synthesized datasets compared to the real dataset, for
    varying UE population.}
    \label{fig:scalability}
\end{figure*}

\parbi{Event breakdown}
%
%
\autoref{tab:macropscopic} compares the differences in event breakdown between
the real and the synthesized datasets of each generator.
Compared with SMM-1 and SMM-20k, \name achieves comparable or smaller
discrepancies across the six event types, \ie within 0.66\%, 2.15\%, and
3.62\% for the three device types respectively,
while SMM-1/SMM-20k has discrepancies up to 1.13\%/1.12\%, 6.17\%/5.92\%, and
3.52\%/3.03\%.
%
When compared to NetShare, \name achieves much smaller discrepancies with the
real dataset for \texttt{ATCH} and \texttt{DTCH}, within 0.02\%, 0.18\%, and
0.08\%, while NetShare exhibits differences up to 0.09\%, 0.53\%, and 0.70\%,
for the three device types respectively.
For the remaining event types, \name and NetShare achieve similar levels of
accuracy.
%

In summary, in terms of event breakdown, \name is capable of synthesizing event
distributions with equal or better accuracy than other generators despite not
requiring any domain knowledge.


\cut{
\parbi{Interarrival time}
Across all device types, SMM-1 generates the most accurate interarrival times,
with maximum y-distances of 1.1\%, 6.4\%, and 2.5\% respectively.
This is followed closely by \name, which gives maximum y-distances of 8.1\%,
9.2\% and 6.3\% respectively.
In contrast, NetShare results in much worse interarrival times, with maximum
y-distances of 24.3\%, 27.8\%, and 32.4\% respectively.

A detailed analysis in \autoref{fig:cdfs_eval_sec} show that SMM-1 and \name
deliver accurate interarrival time distributions, which NetShare fails to.
%
In summary, while transformer-based \name may lag behind domain knowledge-based
SMM-1 in terms of interarrival time, it consistently shows superior performance
when compared with an LSTM-based model, reducing the max y-distance of
interarrival time CDF by 3.02$\times$, 3.01$\times$, and 5.16$\times$ for the
three device types.
}


\subsection{Sensitivity Analysis and Ablation Study}
\label{subsec:ablation_and_sensitivity}

\begin{table}[]
\caption{{Fidelity of \name varying the loss weights of different fields,
and predicting the interarrival time directly instead of its distribution.}}
\label{tab:ablation_and_sensitivity}
\resizebox{\linewidth}{!}{
\begin{tabular}{ll|c|ccc|c}
\toprule
 &  & \multirow{2}{*}{\begin{tabular}[c]{@{}c@{}}Ours\\ (1:1:1)\end{tabular}} & \multicolumn{3}{c|}{\begin{tabular}[c]{@{}c@{}}Varying loss weights\\ (event : arrival : stop\_flag)\end{tabular}} & \multirow{2}{*}{\begin{tabular}[c]{@{}c@{}}No dist.\\ pred.\end{tabular}} \\
 &  &  & 3:1:1 & 1:3:1 & 1:1:3 &  \\ \midrule
\multicolumn{1}{l|}{Violation} & Events & 0.04‰ & 0.04‰ & 0.20‰ & 0.48‰ & 0.10‰ \\
\multicolumn{1}{l|}{} & Streams & 0.2\% & 0.2\% & 0.8\% & 0.4\% & 0.5\% \\ \midrule
\multicolumn{1}{l|}{\multirow{3}{*}{\begin{tabular}[c]{@{}l@{}}Max y\\ distance\end{tabular}}} & Sojourn (\texttt{CONN}) & 6.4\% & 8.4\% & 9.1\% & 6.7\% & 60.8\% \\
\multicolumn{1}{l|}{} & Sojourn (\texttt{IDLE}) & 12.0\% & 11.8\% & 9.3\% & 10.3\% & 75.4\% \\
\multicolumn{1}{l|}{} & Flow length & 3.8\% & 5.0\% & 2.4\% & 3.5\% & 69.9\% \\ \midrule
\multicolumn{2}{l|}{Avg. breakdown diff} & 0.7\% & 0.4\% & 0.2\% & 0.2\% & 0.3\% \\ \bottomrule
\end{tabular}
}
\end{table}

\parbi{Varying Loss Weights}
Up to now, \name is trained to minimize the summation of the losses of the
three fields (event type, interarrival time, and stop flag) with equal
weights.
Next, we study how \name reacts to varying weights used in the summation of the
total loss. We increase the weight of each field by 3x, while keeping the
weights of the other two fields unchanged.
For comparison, we show again the results for \name trained with equal weights
(1:1:1).
\autoref{tab:ablation_and_sensitivity} (middle column) shows that varying the
weights results in little variations for all fidelity metrics.
%
%
{
For example, across all weight combinations, the maximum y-distance of the
sojourn time CDFs varies from 6.4\% to 9.1\% for the \texttt{CONNECTED} state and
9.3\% to 12.0\% for the \texttt{IDLE} state.
}
This shows the training of \name is insensitive to the specific weights chosen.

\parbi{Disabling the Prediction of Distribution Parameters}
We next study \name without design insight 2 (predict distribution parameters
for numerical fields). To this end, we modify \name to output a single scalar
representing the interarrival time, instead of predicting its mean and
variance. Consequently, during generation, there is no random sampling from the
predicted distribution performed.
\autoref{tab:ablation_and_sensitivity} (right column) shows that doing so
impacts fidelity metrics significantly.
While the fidelity in terms of event breakdown remains
unaffected, the maximum y-distances of flow length and sojourn time
distributions increase sharply. For instance, the maximum y-distance of the
flow length CDF increased by 15x, from 3.8\% to 69.9\%.
This shows that predicting the interarrival time distribution parameters is
critical in ensuring \name's fidelity.

\subsection{Scalability Study}

To understand whether \name is scalable in generating datasets with arbitrary
sizes with high fidelity, we run \name inference for varying numbers of times
to generate synthesized datasets containing 10k, 20k, 40k, 80k, and 160k
UEs, respectively.
For each synthesized dataset, we compare it with a subset of the real dataset
of the same size that is randomly sampled from the full testing dataset, which contains
380k UEs.
As shown in \autoref{fig:scalability}, the size of the synthesized dataset has
minimal influence on all the fidelity metrics. This shows that \name is capable
of generating datasets of arbitrary sizes with high fidelity.

\subsection{Adapting to Data Drifts}
\label{subsec:adapting_to_data_drifts}

\begin{table}[]
\caption{Training time in minutes w/ and w/o transfer leraning on an NVIDIA
A100 GPU.}
\label{tab:training_time}
\resizebox{0.8\columnwidth}{!}{
\begin{tabular}{cc|cc}
\toprule
\multicolumn{1}{l}{} & \multicolumn{1}{l|}{} & NetShare & \name \\ \midrule
\multicolumn{2}{c|}{No transfer learning} & 108.36 & 104.40 \\ \hline
\multicolumn{1}{c|}{\multirow{3}{*}{\begin{tabular}[c]{@{}c@{}}\\Transfer\\ learning\end{tabular}}} & First hour & 43.08 & 21.81 \\ \cline{2-4} 
\multicolumn{1}{c|}{} & \begin{tabular}[c]{@{}c@{}}Finetune to each\\ subsequent hour (avg.)\end{tabular} & 30.41 & 9.06 \\ \cline{2-4} 
\multicolumn{1}{c|}{} & Total & 195.12 & 67.12 \\ \bottomrule
\end{tabular}
}
\end{table}

\begin{table}[tp]
\caption{Maximum CDF y-distances w/ and w/o transfer learning.}
\label{tab:xfer_learning_fidelity}
\resizebox{\columnwidth}{!}{
\begin{tabular}{ll|cc|cc}
\toprule
 &  & \multicolumn{2}{c|}{w/o xfer learning} & \multicolumn{2}{c}{w/ xfer learning} \\
 &  & NetShare & CPT-GPT & NetShare & CPT-GPT \\ \midrule
\multicolumn{1}{l|}{\multirow{2}{*}{Violation}} & Events & 2.78\% & 0.07\% & 3.39\% & 0.05\% \\
\multicolumn{1}{l|}{} & Streams & 34.58\% & 0.40\% & 37.57\% & 1.00\% \\ \midrule
\multicolumn{1}{l|}{\multirow{3}{*}{\begin{tabular}[c]{@{}l@{}}Max y\\ distance\end{tabular}}} & Sojourn (\texttt{CONN}) & 36.28\% & 9.39\% & 13.21\% & 12.48\% \\
\multicolumn{1}{l|}{} & Sojourn (\texttt{IDLE}) & 21.16\% & 13.40\% & 28.43\% & 8.98\% \\
\multicolumn{1}{l|}{} & Flow length & 3.30\% & 7.32\% & 2.24\% & 3.08\% \\ \bottomrule
\end{tabular}
}
\end{table}

Next, we measure the training time needed to train each model to
capture data drift across different hours of the day.
To synthesize traffic across multiple hours, the operator
can either train a single model on the entire dataset spanning over multiple hours,
or train multiple specialized models, each tailored to a specific hour.
In the latter setup, the operator may train one specialized model from scratch,
and subsequently perform transfer learning on it to generate the remaining
models. We measure the training time under both setups.

Since the traditional training loss for GAN does not necessarily correlate with
the quality of the generated samples~\cite{lucic2018gans,mescheder2018training}
\footnote{
Generator and discriminator losses can oscillate, and a decrease in loss is not
a reliable indicator of improved generator performance. To assess whether
training has converged, it is more effective to evaluate the models at various
epochs throughout the training process.
},
we devise the following heuristics to compare the training time of NetShare and
\name in a fair manner:
%
%
For each model, we save checkpoints every $N$ epochs, producing $K/N$
checkpoints in total, where $K$ is the model's total number of epochs.
For each checkpoint, we generate synthetic datasets and compute the fidelity
metrics against the validation set.
We then rank the checkpointed models for each fidelity metric, ranging from 1 (best) to
\textit{N} (worst). Finally, we sum the rankings for each checkpoint across
fidelity metrics, select the top 20\% checkpoints with the smallest ranking
sums, and pick the earliest checkpoint among them.

\autoref{tab:training_time} shows the training time for a single model on
6-hour traces, or six specialized models through transfer learning.
Without transfer learning, NetShare and \name require a similar time to train a
model on 6-hour traces, taking 108.36 and 104.40 minutes, respectively.
Employing transfer learning, \name shows significantly reduced training time
for adapting a 1-hour model to another hour, requiring only 9.06 minutes,
much shorter than NetShare's 30.41 minutes.
This difference may be attributed to \name's use of supervised training, as
opposed to NetShare's GAN training which is known to be difficult to
converge~\cite{kodali2017convergence,thanh2020catastrophic,srivastava2017veegan}.
Consequently, it takes merely 67.12 minutes to generate all six \name models each
for a specific hour. In comparison, NetShare cannot benefit from transfer
learning, taking 195.12 minutes in total.

\autoref{tab:xfer_learning_fidelity} shows the fidelity metrics for the trace
generated for the 4-th hour, with and without transfer learning. We observe
 that the use of transfer learning does not have an obvious impact on accuracy
for either NetShare or \name. 
For instance, when trained with transfer learning, NetShare shows better
\texttt{CONNECTED} state sojourn time and flow length distributions, but worse
semantic violations and \texttt{IDLE} state sojourn times.
For \name, employing transfer learning results in better \texttt{IDLE} state
sojourn times and flow length distributions, but worse \texttt{CONNECTED} state
sojourn times and slightly more state violations.

In summary, \name demonstrates much higher training efficiency on transfer
learning, resulting in substantially reduced training time to synthesize
multi-hour traces, with minimal impact on fidelity.

\subsection{Data Memorization}
Ideally, the generated dataset should closely mirror the real dataset in terms
of aggregated statistical behavior, yet maintains diversity, \ie the
generator should not memorize or replicate the training dataset.
Achieving this is crucial for ensuring that downstream applications are
exposed to a wide variety of traffic behavior, while also preventing 
leakage of private information.

\parbi{Methodology}
We adopt the methodology from the natural language
domain~\cite{mccoy2023much,carlini2022quantifying,yang2024unveiling} to
quantify data memorization in \name.
Specifically, we extract all \textit{n}-grams from both the generated and real
dataset, where an \textit{n}-gram is a continuous subsequence of length
\textit{n}.
For each \textit{n}-gram in the generated dataset, we scan through the
\textit{n}-grams in the training set and check for repeats. 
We report the percentage of \textit{n}-grams from the generated dataset with at
least one repeat found.

Unlike natural languages, cellular control-plane traffic is multi-modal,
consisting of categorical (event type) and numerical (interarrival time)
fields.
We consider two \textit{n}-grams to repeat if they share the same sequence of
event types, and every corresponding pair of interarrival times fall within a
relative tolerance $\epsilon$, \ie $(1-\epsilon) < t_{generated,i} / t_{real,i}
< (1+\epsilon)$, for $i=1,2,..,n$.

Generally, a larger $n$ reduces the likelihood of repetitions as longer
sequences are less likely to repeat, while a larger $\epsilon$ increases the
likelihood of repetitions.
We examine various ranges of $n$ and $\epsilon$.

\begin{table}[]
\caption{Percentage of \textit{n}-grams in the generated dataset that repeats
from the training dataset.}
\label{tab:memorization}
\begin{tabular}{c|cc}
\toprule
 & $\epsilon$=10\% & $\epsilon$=20\% \\ \midrule
n=5 & 57.879\% & 80.305\% \\
n=10 & 0.003\% & 0.287\% \\
n=20 & 0.000\% & 0.000\% \\ \bottomrule
\end{tabular}
\end{table}

\parbi{Results}
\autoref{tab:memorization} shows the percentage of \textit{n}-grams in the
generated dataset that are repeated from the training dataset, for UEs of
phone device type.
We make the following observations:
(1) Short sequences of length 5 have a high likelihood of repetition.
However, such very short repetitions should not be considered memorization, as
they are often constrained by control-plane protocols, rather than driven by
end user behaviors.
For example, \texttt{HO} is always followed by \texttt{TAO} in the
\texttt{CONNECTED} state, and consecutively alternating \texttt{SRV\_REQ} and
\texttt{S1\_CONN\_REL} are common.
(2) Sub-sequences longer than 10 are rarely repeated, even under loose
tolerances of $\epsilon=10\%$ and $20\%$.
For example, with $n=10$, only 0.003\% and 0.287\% of \textit{n}-grams are
found to be repetitions under 10\% and 20\% tolerances; with $n=20$, no
repeating \textit{n}-grams are found under either tolerance.

In summary, \name learns generalized information from the training set,
instead of memorizing and repeating any samples from the training set.

\section{Related Works}

\noindent
{\bf Traditional Modeling-based Traffic Generators.} 
There have been a large body of work on using statistical models to 
model and generate traffic, \eg~\cite{rolland2007litgen, sommers2004harpoon, weigle2006tmix, vishwanath2009swing}.
However, these works rely on observed statistics from Internet traffic,
which do not apply to control-plane traffic of cellular networks~\cite{meng2023modeling}.
For the control-plane traffic of mobile networks, \cite{dababneh2015data} 
models the total traffic volume instead of interarrival times and the dependence between events.
The state-of-the-art control traffic generator, SMM~\cite{meng2023modeling},
requires domain knowledge and suffers from its model complexity as discussed in~\autoref{subsec:smm}.
 
\noindent
{\bf ML-based Generators for Data Traffic.}
Many works study modeling Internet traffic using ML-based approaches.
STAN~\cite{xu2020stan} uses autoregressive neural models to synthesize
flow-level traffic but fails to capture fine-grained features such as packet
interarrival times.
The GAN-based solutions (\eg~\cite{gulrajani2017improved, wang2020packetcgan,
xu2019modeling, doppelgan:imc2020}) ignore temporal aspects of the traffic and
are shown to have suboptimal fidelity compared with
NetShare~\cite{netshare:sigcomm2022}.
A few recent works~\cite{jiang2024netdiffusion,sivaroopan2024netdiffus} model
each pcap flow into a 2D image, and employ stable diffusion models to generate
synthetic images and convert them to pacp flows.
However, \cite{jiang2024netdiffusion} is only capable of generate
fixed-length flows, and requires domain knowledge to post-process the generate
traffic to maintain high fidelity, whereas \cite{sivaroopan2024netdiffus} can
only generate single-modality traffic.


\section{Conclusions and Future Work}

We proposed a transformer-based traffic generator framework for
cellular network control-plane traffic critically needed in research
and innovation on mobile core network design and implementation. \name
is based on a key observation that the generative model requires no domain knowledge
and its attention mechanism 
has the potential to capture complex dependencies among the stream of control
events by each UE.
Our evaluation shows \name synthesizes control-plane traffic with
comparable fidelity as prior-art SMM but without domain knowledge,
significantly higher fidelity than the state-of-the-art GAN-based
scheme in terms of stateful semantics and interarrival time of control
events, and does not memorize streams from training traffic.
%

Due to the lack of systematic support for 5G trace collection, in this
paper, we could only use LTE's trace to showcase our transformer-based approach.
The versatility of CPT-GPT from not relying on domain knowledge
makes it generally applicable to synthesizing control-plane traffic
in a wide variety of scenarios.
In future work, we plan to evaluate
\name for next-generation networks (5G, 6G, et al.)
and complex scenarios such as mixed LTE/5G networks
involving frequent inter-RAT handovers.
%
Additionally, given the coexistence of 4G and 5G in current
deployments~\cite{k2024unveiling}, we intend to evaluate \name's performance on
4G/5G co-existing traffic, once such datasets become available.
%
Finally, we plan to evaluate \name's effectiveness on downstream applications
as such applications become publicly available in the future.

%

\begin{acks}
We thank the anonymous reviewers and our shepherd Chase Jiang for their helpful
comments. This work is supported in part by NSF grant CNS-2312834.
\end{acks}
\newpage

\bibliographystyle{ACM-Reference-Format}
\balance
\bibliography{reference}

\appendix
\section*{APPENDIX}

\section{\bf Ethics}
%
Although generative ML modes can potentially memorize individual records in
training~\cite{hayes2017logan}, \name is trained on LTE data with UE-specific
information obfuscated, and therefore both the trace used for training and the
trace it synthesizes do not reveal UE-specific information. Hence this
work raises no ethical concerns.

\section{\bf Additional Dataset Statistics}
\label{sec:additional_dataset_statistics}

\begin{figure}[H]
    \includegraphics[width=0.8\columnwidth]{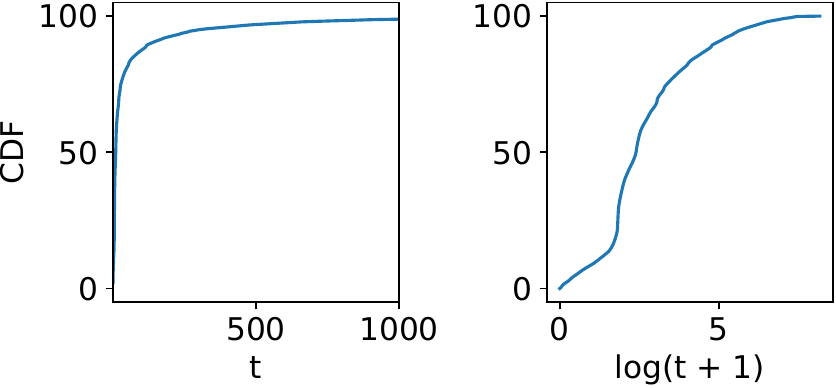}
    \caption{Interarrival time ($t$, unit in seconds) distribution for UEs of device type phone.}
    \label{fig:interarrival_cdf_phone}
\end{figure}

\autoref{fig:interarrival_cdf_phone} (left figure) shows the interarrival time
distribution for UEs of the phone device type.
The distribution exhibits a long-tailed pattern, characterized by a higher frequency of short interarrival times and a lower frequency of long ones.
Therefore, \name applies log transformation to the interarrival times,
effectively reducing the impact of the long-tailed distribution, as shown in
the right figure.

\end{document}